# Unsteady Land-Sea Breeze Circulations in the Presence of a Synoptic Pressure Forcing

Mohammad Allouche[1,2], Elie Bou-Zeid[1*], and Juho Iipponen[3]

[1]Department of Civil and Environmental Engineering, Princeton University, New Jersey

[2]Lawrence Livermore National Laboratory, Livermore, California

[3]Program in Atmospheric and Oceanic Sciences, Princeton University, New Jersey

[*]Corresponding author: Elie Bou-Zeid (ebouzeid@princeton.edu)

**Key Points:**

- Unsteady large-eddy simulations of land-sea breezes reveal asymmetric dynamics when the geostrophic wind is oriented from sea to land, versus land to sea.

- The diurnal dynamics result in non-equilibrium flows , and the sea patch is found to control the final turbulence-mean flow equilibrium response.

- An autonomous clustering, to delineate these macro circulations into four regimes (canonical, transitional, shallow land-driven, and advected), is successfully tested.

**Abstract**

Unsteady land-sea breezes (LSBs) resulting from time-varying surface thermal contrasts $\Delta\theta(t)$ are explored in the presence of a constant synoptic pressure forcing, $M_g$, when the latter is oriented from sea to land ($\alpha=0°$), versus land to sea ($\alpha=180°$). Large eddy simulations reveal the development of four distinctive regimes depending on the joint interaction between ($M_g$, $\alpha$) and $\Delta\theta(t)$ in modulating the fine-scale dynamics. Time lags, computed as the shifts that maximize correlation coefficients of the dynamics between transient and the corresponding steady state scenarios at $\Delta\theta=\Delta\theta_{max}$, are found to be significant and to extend 2 hours longer for $\alpha=0°$ compared to $\alpha=180°$. These diurnal dynamics result in non-equilibrium flows that behave differently over the two patches for both $\alpha$'s. Turbulence is found to be out of equilibrium with the mean flow, and the mean itself is found to be out of equilibrium with the thermal forcing. The sea surface heat flux is consistently more sensitive than its land counterpart to the time-varying external forcing $\Delta\theta(t)$, and more so for synoptic forcing from land-to-sea ($\alpha=180°$). Hence, although the land reaches equilibrium faster, the sea patch is found to exert a stronger control on the final turbulence-mean flow equilibrium response. Finally, vertical velocity profile at the shore and shore-normal velocity transects at the first grid level are shown to encode the multiscale regimes of the LSBs evolution, and can thus be used to identify these regimes using $k$-means clustering.

**Plain Language Summary**

Advancing our understanding of atmospheric circulation dynamics in coastal regions is essential as these zones host most of the world's population and economic activity. These dynamics have an oversized influence on our ability to better predict pollution ventilation and urban heat island induced-dynamics in coastal cities. In this study, we investigate the diurnal cycles of land-sea breeze circulations, with the aid of a state of art numerical simulation tool in the presence of increasingly large scale weather systems. These simulations depict added complexities as the land-sea surface thermal contrasts, mean flow, and turbulence are found to be consistently out of equilibrium. The gained physical insights from this analysis offer new explanations on why coarse weather and climate models inherently fail to reproduce such unresolved processes, but also offer a pathway for better understanding and prediction. Specifically, an autonomous clustering approach to independently classify these multiscale circulations is successfuly tested against visual categorization, identifying four emerging regimes.

## 1 Introduction

Real-world land-sea breezes (LSBs) are perceptibly unsteady. Simulating such transient scenarios is thus motivated by the need to provide a more realistic picture of the evolution of winds within the coastal Atmospheric Boundary Layer (ABL), and how they respond to the interaction of the LSBs with a synoptic pressure forcing. This would be beneficial in many applications including (i) advancing our knowledge on the diurnal potential of offshore wind power where these LSBs might amplify or reduce the available power (Howland et al., 2020; Kumer et al., 2016; Porté-Agel et al., 2020), (ii) improving the forecast of pollutant dispersion (Levy et al., 2009; Llaguno-Munitxa & Bou-Zeid, 2018; Lyons, 1995), as well as rainfall and cloud formation (Mazon & Pino, 2015) in coastal regions; and (iii) characterizing the implications of LSBs on the environment and the microclimatic conditions of coastal cities (Bauer, 2020; Lin et al., 2008; Lo et al., 2006). In addition, from energy systems and design perspectives, forecasting the intermittent wind power is a key element to ensure a stable electricity grid, analyze wind turbine blades fatigue, and optimize hub height and wind farm layouts (Díaz & Guedes Soares, 2020).

Previous studies on the topic of transient land-sea breezes have used observations and simulations (Antonelli & Rotunno, 2007; Atkison, 1995; Cana et al., 2020; Porson et al., 2007; Rizza et al., 2015; Segal et al., 1997; Sills et al., 2011; Steyn, 2003; Yang, 1991) to identify a number of important parameters that control the dynamics. These include: (i) the transient difference $\Delta\theta(t)$ between land surface temperature $\theta_L$ and sea surface temperature $\theta_S$ and its natural time scale (usually 24 hours), (ii) the speed ($M_g$) and direction ($\alpha$) of the synoptic pressure gradient forcing, (iii) the latitude as it influences the Coriolis parameter $f$ and the associated time scale of inertial oscillations, (iv) land and sea roughness lengths for momentum and for heat, (v) land topography, and (vi) the height of the inversion and its strength. This vast parameter space prompted the authors of this study to propose a reduction to a set of non-dimensional parameters (Allouche, Bou-Zeid, et al., 2023) to facilitate generalization of the physical findings and to guide numerical and experimental studies. In that prequel paper, we focused on analyzing the competing effects of synoptic forcing and thermal contrast under steady state conditions, identifying several distinct regimes of the LSBs and notable asymmetry in the synoptic effect when the geostrophic wind blew from land-to-sea or vice versa. The important transient effects were not considered, but they are the focus of the present study.

Previous similar studies on the topic of transient land-sea breezes used two-dimensional numerical models and compared to laboratory experiments. Sha et al., 1991 investigated how the Kelvin-Helmholtz (KH) instability affects the diurnal structures of these LSBs under light wind sonditions. Similarly, Yoshikado, (1992) examined the basic characteristics of the daytime heat-island circulation and its interaction with the sea breeze and concluded that the urban heat-island effect could delay the inland transport of urban pollutants and potentially prevent their dispersion. Using a three-dimensional cloud resolving model, Dailey and Fovell (1999) explored how the horizontal convective rolls (HCRs) can modulate the convection process along the presence of sea-breezes (SBs) along $α=0°$. Similarly, Fovell and Dailey (2001) examined again the overall convective activity, with an alongshore ambient flow this time ($α=90°$). Jiang et al., (2017) studied, using large-eddy simulation, how the streaky turbulent structures modulate SB front characteristics over urban-like coasts under strong wind shear and moderate buoyancy i.e., rendering the SB front into three-dimensional structures with strengthened updrafts. Also, Fu et al., (2021), with the aid of large-eddy simulation, explored over a peninsula the different generations of deep-convection initiation through the collision of two sea-breeze fronts. They reported how a decreased land heat flux weakens each of the identified generation until one reaches shallow convection. Others conducted laboratory experiments to analyze flow patterns associated with the land and see breezes (Mitsumoto et al., 1983; Moroni & Cenedese, 2015; van der Wiel et al., 2017).

The enormous differences in heat capacity and available energy partitioning to latent heat between the sea (large) and land (small) explain why water surface temperature changes reduce to near-zero during a diurnal cycle. In contrast, land surface warms and cools more intensely due to its lower thermal conductivity and heat capacity (and thus lower thermal admittance or effusivity), as well as higher Bowen ratio. Consequently, local meteorology is here largely controlled by the diurnal cycle of land surface temperature, along with the synoptic scale forcing. When synoptic pressure forcings are added, they induce a broad range of spatial transitions in buoyancy fluxes, jointly modulating these LSBs with the thermal contrast. One expected feature of such circulations is non-equilibrium turbulence exhibited as a hysteretic response of the surface kinematic fluxes, mean flow, and turbulence relative to the imposed forcing. Since most geophysical parameterizations still assume turbulence-mean equilibrium (Bou-Zeid et al., 2020a; Huang et al., 2013; Mahrt & Bou-Zeid, 2020), non-equilibrium could offer a partial explanation as to why climate models are still showing consistent upwelling biases (Lembo et al., 2019; Richter-Menge et al., 2017) (in addition to the systematic winds biases that are also attributed to the disparate resolutions of the atmospheric and oceanic grids. Therefore, an analysis with a prescribed land surface temperature pattern that reflects a canonical diurnal variation would offer new insights into the non-linear interplay between various mechanisms that evolve at different time scales and rates including (i) the inherent thermal and kinetic inertia (memory) effects of the air, (ii) the evolution of turbulent structures associated with the diurnal land surface forcing (Salesky et al., 2017), (iii) sea breeze (SB) and land breeze (LB) tendency for initiation, (iv) the strength of the advected inflows and their inherent asymmetric nature depending on the angle $\alpha$ between the shore and geostrophic wind.

In this paper, we consider the same numerical setup introduced in (Allouche, Bou-Zeid, et al., 2023) for steady thermal contrasts, but here we account for the transient diurnal cycle of the surface temperature contrasts between land and sea $\Delta\theta(t) = \theta_L(t) - \theta_S$. As the expression suggests, we impose a constant sea surface temperature ($\theta_S$) and a time-varying land surface temperature ($\theta_L(t)$), while also including a constant synoptic pressure forcing. We are thus able to compare the unsteady simulations we conduct here to the reference steady state scenarios we reported before (Allouche, Bou-Zeid, et al., 2023), hereafter referred to by ABI23, where the surface thermal contrast is fixed at $\Delta\theta = 10$ K. Specifically, we aim to answer the following questions: (Q1) How do the joint interaction between the synoptic variables ($M_g$, $\alpha$) and transient surface thermal contrasts $\Delta\theta(t)$ modulate the fine-scale dynamics and structures of the LSBs? The complexity of the emergent circulations and the disparate flow features over the two patches motivate the second question: (Q2) How do the surface kinematic fluxes, shoreline mass exchanges, and turbulence evolve with the external parameters ($M_g$, $\alpha$) and $\Delta\theta(t)$ and is there hysteresis? Finally, visual identification of the circulation regimes motivates the last question: (Q3) What are the best criteria to autonomously delineate the emerging LSBs from a macro (large scale) and micro (near surface) perspectives?

In section 2, an expanded dimensional analysis, based on the one outlined in ABI23, is presented. In section 3, the numerical experiments design (boundary and initial conditions) is detailed. The thermal circulation flow structures are then discerned and visually categorized in sections 4 and 5 (answering Q1). In section 6, land and sea surface kinematic fluxes, shoreline exchanges, and turbulence are investigated (answering Q2). In section 7, criteria are proposed to demarcate these LSBs from a macro and micro categorization (answering Q3). Finally, conclusions and implications of this work are drawn in section 8.

## 2 Dimensional analysis

This section builds on the dimensional analysis outlined by ABI23, where a set of dimensionless groups was constructed based on Buckingham's Pi theorem for the steady state dynamics where that theorem applies. The pertinent dimensional external input parameters to our problem involve three length scales: (i) $z_{0,L} = 0.002$ (m), the land surface roughness length, (ii) $z_{0,S} = 0.002$ (m), the sea surface roughness length (here not a function of wind), and (iii) $z_i = 1600$ (m), the ABL height. The thermal roughness lengths are assumed to be fully determined by, but not necessarily equal to, their momentum counterparts. In addition, two velocity scales of are of relevance: (i) $M_g = [U_g^2 + V_g^2]^{1/2}$ the geostrophic wind magnitude, and (ii) $W_*(t) = [g\beta z_i(\Delta\theta(t))]^{1/2}$, a time varying convective buoyant velocity scale where $g = 9.81$ (m s$^{-2}$) is the gravitational acceleration, $\beta = 1/\theta_r$ (K$^{-1}$) the volumetric expansion coefficient of air, $\theta_r = 300$ (K) a reference temperature in the Boussinesq approximation sense, and $\Delta\theta(t) = \theta_L(t) - \theta_S$ (K), (refer to Fig. 1, top).

For unsteady state analysis, the flow physics are also influenced by two time scales arising from (i) $2\pi/f_c \approx 12.5$ h, the inertial period, where $f_c = 1.394\times10^{-4}$ Hz is the Coriolis parameter (at lower latitudes, this time scale is much longer and less important, reflecting the waning influence of the Coriois force), and (ii) $2\pi/\omega = 24$ h, the diurnal earth surface heating/cooling rate (corresponding to the variability in buoyancy forcing, $\tau_{forcing} = 24$ h), shown later to dominate the transient dynamics. The kinetic and thermal inertia times scales will also modulate the response. These time scales will depend on the problem setup, as well as on the extent of the thermal circulation, but should be much shorter than 12.5 or 24 h. We will revisit these scales when we discuss the hysteris of the dynamics. Finally, the alignment angle $\alpha$ between the shore and the geostrophic velocity vector is the only non-dimensional variable that forms one of the dimensionless groups. Therefore, any output of the numerical experiments, with $n=8$ input dimensional parameters and $k=2$ independent dimensions (L, T), can be also non-dimensionalized and expressed in terms of $n-k=6$ dimensionless groups. The five identified $\Pi$'s by ABI23, some of which are here variable in time, are stated here for reference: two dimensionless parameters are related to the introduced length scales $\Pi_1 = z_{0,L}/z_{0,S}$ (surface roughness contrast expressed as inner scales ratio) and $\Pi_2 = z_i/z_{0,S}$ (outer to inner scales ratio, commensurate with a Reynolds number of the problem). Another dimensionless group is associated with the velocity scales $\Pi_3(t) = Ri^{-1}(t) = M_g^2/[g\beta z_i(\theta_L(t)-\theta_S)] = M_g^2/W_*(t)^2$, which is an inverse bulk Richardson number that relates inertia to buoyancy. An inverse convective Rossby number also arises, comparing the convective/buoyant and inertial time scales $\Pi_4(t) = (z_i/W_*(t))/(2\pi/f_c)$, and $\Pi_5 = \alpha$. The last new emerging number, resulting from the aforementioned two time scales ($2\pi/f_c \approx 12.5$ h and $\tau_{forcing} = 24$ h), is $\Pi_6 = 2\pi/f_c / \tau_{forcing} \approx 0.52$. The time evolution can then also be described as a function of the non-dimensional time $\omega t$. The time variability of the problem precludes exact applicability of the Buckongham Pi theorem, and thus exact similarity. But the non-dimensionalization remains an effective tool for reducing the parameter space and elucidating the physics.

In this paper, the simulations consider only the effects of the synoptic variables $M_g$ (m/s), which here takes on the value 0, 0.4, 1.2 and 2 ms$^{-1}$, $\alpha = 0°$ or 180°), and $\Delta\theta(t) = \theta_L(t) - \theta_S$. $\Pi_1$ is set to 1 to isolate the impact of the thermal torque (due to temperature differences) from that of the stress torque (that is mainly controlled by roughness differences) (Bou-Zeid et al., 2020b). In addition, $\Pi_2 = 8\times10^5$ is fixed to a high value ensuring a high Reynolds number and large separation of turbulent to friction scales. The convective Rossby number $\Pi_4$ must vary in time since it is a function of the thermal contrast.

## 3 Large eddy simulations

### 3.1 Governing equations

As described by ABI23, the used LES code solves the spatially filtered incompressible mass continuity and Navier-Stokes momentum equations using the Boussinesq approximation for the mean state, in addition to the advection-diffusion equation for the potential temperature, which are given respectively as follows:

$$\frac{\partial \tilde{u}_i}{\partial x_i} = 0, \tag{1}$$

$$\frac{\partial \tilde{u}_i}{\partial t} + \tilde{u}_j \left( \frac{\partial \tilde{u}_i}{\partial x_j} - \frac{\partial \tilde{u}_j}{\partial x_i} \right) = -\frac{1}{\rho}\frac{\partial \tilde{p}^*}{\partial x_i} - \frac{\partial \tau_{ij}}{\partial x_j} - g\frac{\tilde{\theta}'}{\theta_r}\delta_{i3} + f_c(U_g - \tilde{u}_1)\delta_{i2} - f_c(V_g - \tilde{u}_2)\delta_{i1}, \tag{2}$$

$$\frac{\partial \tilde{\theta}}{\partial t} + \tilde{u}_j \frac{\partial \tilde{\theta}}{\partial x_j} = -\frac{\partial \pi_j}{\partial x_j}. \tag{3}$$

The tilde (~) represents filtered quantities (omitted throughout the rest of paper for simplicity since we only deal with filtered LES outputs); $x_i$ (or $X_i$) is the position vector; $\tilde{u}_i$ (or $U_i$) is the resolved velocity vector (the indices $i$ and $j$ span the three directions 1 for $X$ (across shore), 2 for $Y$ (along shore), and 3 for $Z$ (vertical)); $t$ is time; $\rho$ is the density of air; $p^*$ is a modified perturbation pressure that includes the resolved and unresolved turbulent kinetic energy; $\tau_{ij}$ is the anisotropic part of the full sub-grid scale (SGS) stress tensor $\sigma_{ij} = \widetilde{u_i u_j} - \tilde{u}_i \tilde{u}_j$; $\theta$ is the potential temperature with its reference value $\theta_r$; $\theta'$ is the deviation of the local $\theta$ from its horizontal planar average in the LES; $\delta_{ij}$ is the Kronecker delta; and $\pi_j = \widetilde{\theta u_j} - \tilde{\theta}\tilde{u}_j$ is the SGS heat flux vector. The prime always denotes the turbulent perturbation from a mean state, indicated by an overbar. The SGS stress and heat flux are parameterized using a scale-dependent Lagrangian dynamic model that was extensively validated (Bou-Zeid et al., 2005; Huang & Bou-Zeid, 2013; Kumar et al., 2006), with a constant SGS Prandtl number of 0.4. The computational grid is a uniform structured mesh, staggered in the vertical direction to compute vertical derivatives using second-order centred differences. Horizontal derivatives are computed spectrally, and thus the domain must be periodic in the horizontal directions, but we modify it to impose inflow in the $X$ direction, as clarified in subsection 3.3. An Adams-Bashforth second-order explicit time advancement scheme is used.

### 3.2 Suite of simulations

The designed simulations, illustrated in Fig. 2, span seven cases, one case for $M_g$=0 and six cases with all combinations of {$M_g$ (m/s): 0.4, 1.2, 2 and $\alpha$: 0°, 180°}. The domain size is $L_x$ = 80 km × $L_y$ = 5 km × $L_z$ = $z_i$ = 1.6 km. The corresponding baseline number of grid points ($N_x$, $N_y$, $N_z$) is 384×24×64. The grid resolution ($dX = dY$ = 208 m, $dZ$ = 25m) is course relative to typical ABL LES studies, and the simulations thus represent a very large eddy simulation of the problem, where a significant fraction of the production range may not be resolved (Pope, 2001), but the analyses focus primarily on the mean flow and secondary circualtions that are well resolved.

### 3.3 Boundary and initial conditions

For both scenarios (null or positive $M_g$), the boundary conditions are set periodic in the $Y$ direction for velocities and temperature in all simulations to mimic an infinite coastline. A stress-free impermeable lid boundary condition at top of the domain is imposed for the velocities because any resultant LSB might be sensitive to the free tropospheric profiles and inversion strength (Cioni & Hohenegger, 2018), and in this paper we want to avoid adding yet another parameter to the investigation (our setup thus mimics a very strong inversion). The surface temperatures of the two patches are imposed as $\theta_S$ = 278.15 K and ($\theta_L(t) = \theta_S + \Delta\theta_{max}\sin(\omega t)$, where $\Delta\theta_{max} = \theta_{L,max} - \theta_S$ =10 K, $t$ in hours), and surface stresses and heat fluxes are calculated using a local equilibrium wall-model based on a log-law with Monin-Obukhov stability correction (Bou-Zeid et al., 2005; Ghannam & Bou-Zeid, 2021).

As sketched in Fig. 2 of ABI23, two equal buffer regions on the streamwise boundaries of the domain are implemented with tangent hyperbolic interpolation to act as smooth transitions of the streamwise flow (since the numerical code is pseudo-spectral and periodic in $X$) (Lund et al., 1998; Spalart, 1988), and realistic upstream inflows for a range of geostrophic wind conditions (for both $\alpha$ values) are generated using precursor simulations. The length of the buffer area ($L_{Buffer}$) on each side is 1/16 of $L_x$ (≈ 5 km). A region twice the buffer length on either side of the domain in *X* is excluded from analysis to further minimize the impact of streamwise boundary conditions. As described in ABI23, the *X* boundary conditions are treated differently when (i) $M_g$=0: no-slip boundary condition (NS-BC: $U=V=W=0$) or (ii) $M_g \neq 0$: the inflows for velocities and temperature are generated in precursor simulations with periodic boundary conditions in both *X* and *Y* directions.

The imposed surface temperature (and thermal roughness lengths) in the precursor simulations is set equal to the value of the sea ($\theta_S$ = constant temperature) when *α*=0° or land ($\theta_L(t)$ = time varying temperature) when *α*=180° to represent an infinitely homogeneous upstream fetch. The nature of these generated inflows is thus drastically different for the two considered directions. As demonstrated in the bottom panel of Fig. 1, for *α*=180°, transiently saved inflows, which mimic the continuous evolution of structures in response to the same sinusoidal diurnal land surface temperature profile, are fed at the same physical time in the heterogeneous simulation. Therefore, such inflows inherently span a wide range of stabilities. On the other hand, for *α*=0°, the generated inflows mimic an infinite persistently near-neutral sea. In this case, the saved inflows are only needed over the last quarter inertial period ($0.25 \times \tau_f$=3.125 h) out of the total sea precursor simulations time ($T$=4.25×$\tau_f$=53.125 h), and these inflows are recycled every $0.25 \times \tau_f$ in the *α*=0° main simulations. Since the internal dynamics of the main domain are evolving, this recycling will not produce identical flows repeated every $0.25 \times \tau_f$ in the main domain.

In all the simulated scenarios in this paper, the initial air temperature in the whole analysis domain is uniformly set equal to the sea surface temperature ($\theta_{init}=\theta_S$=278.15 K). Similarly, the land surface temperature is initiated with $\theta_L(t_0) = \theta_S$ = 278.15 K. All analyzed statistics here (null and finite synoptic forcing) refer to the third diurnal cycle, i.e., the 24-hour period that follows the initial 48-hour simulation spin up, as illustrated in Fig. 1, bottom panel. The same synoptic pressure gradient (as a geostrophic wind) used to generate the inflow is imposed as well in the main simulation domain, and thus we do not rely solely on the inflow inertia to represent the synoptic flow. A schematic diagram of the three-dimensional domain and the boundary conditions is given in Fig. 2.

## 4  Thermal circulation structures: Snapshot dynamics at selected times

The features of the simulated transient LSBs, and how they are jointly modulated by ($M_g$, $\alpha$) and $\Delta\theta(t) = \theta_L(t) - \theta_S$, are explored in this section. Figs. 3 and 4 depict pseudocolor plots of the along-shore and time averaged stream-wise $\langle U \rangle_{y,t}$ (brackets will henceforth denote averaging along-shore in $y$ and in time over 1 hour centered around the $\Delta\theta(t)$ reported in the figure or analysis) through an X-Z slice of the analysis domain [X=10 km-70 km] (twice the buffer zone is excluded). The series of column panels in each figure correspond to the same $\alpha$ with increasing $M_g$, and the series of row panels in each figure correspond to the same ($M_g$, $\alpha$) at different times and thus different contrasts $\Delta\theta(t)$ (subscript 1: corresponds to quarter-periods prior to $\Delta\theta=\Delta\theta_{max}$ and $\Delta\theta=\Delta\theta_{min}$, and subscript 2: corresponds to quarter periods following $\Delta\theta=\Delta\theta_{max}$ and $\Delta\theta=\Delta\theta_{min}$). The selected color bar ranges for $\langle U \rangle_{y,t}$ (m/s) are kept unchanged in all figures to provide a basis for a clear comparison. Pseudocolor plot movies of $\langle U \rangle_{y,t}$ and $\langle \theta \rangle_{y,t}$ are also generated for these scenarios (see movies through the links provided in section 5). First and as depicted in these figures, the four regimes that emerge will be introduced and briefly in the following subsections, regime 1: canonical LSBs (subsections 4.1.1: canonical SB and 4.1.2: canonical LB), regime 2: transitional LSBs (subsection 4.2), regime 3: land-driven SB (subsection 4.3), and regime 4: advected LSBs (subsection 4.4). Second, an overview of the transient behavior in all scenarios with respect to $\Delta\theta(t)$ is discussed afterwards, in section 5.

*4.1  Canonical LSBs*

With a weak or zero $M_g$, a canonical counterclockwise deep thermal circulation (canonical SB) is established because of the positive thermal contrast between the colder sea (left patch) and the warmer daytime land (right patch) when $\Delta\theta >0$. When the sign of this thermal contrast is reversed ($\Delta\theta <0$) for the same setup scenario, the resultant steady state LSB is just a mirror image of the original canonical SB; that is, another canonical clockwise deep thermal circulation (canonical LB) would form.

The canonical sea breeze can be noted in the subplots corresponding to $M_g$=0 at $\Delta\theta_1$=5, $\Delta\theta$=10, $\Delta\theta_2$=5, $\Delta\theta_2$=0 and $\Delta\theta_1$=−5) in Fig. 3. Such regime is featured also for both directions $\alpha$=0° and 180°, but more consistently for $\alpha$=0°, in the subplots that correspond to $\alpha$=0°, $M_g$=0.4 and 1.2: $\Delta\theta_2$=5, $\Delta\theta_2$=0 and $\Delta\theta_1$=−5 of Fig. 3 and $\alpha$=180° with $M_g$=0.4 only and for $\Delta\theta_2$=5, $\Delta\theta_2$=0 of Fig. 4. Canonical land breeze, on the other hand, is shown in the subplot ($M_g$=0, $\Delta\theta_2$=−5), and it prevails upon continuing the cycle in the subplot ($M_g$=0, $\Delta\theta_1$=0). Canonical land breeze LSBs exist also along both directions in the subplots that correspond to ($\alpha$=0° and 180°, $M_g$=0.4, $\Delta\theta$=−10) of Figs. 3 and 4.

*4.2 Transitional thermal circulation*

When the LSB is shifting regimes between the clear canonical reference states, we observe a transitional state with no definite pattern; e.g., refer to the initiation of land breeze in the lower ABL with a canonical SB background in the subplot {$M_g$=0, $\Delta\theta_1$=−10} in Fig. 3. One can recognize here three circulation cells because of the created entrainment zones. Such regimes generally inherit remnants of clearly categorized circulations (reference states), but acquire features of an emerging regime that slowly develops over time. When the synoptic wind blows from sea to land ($\alpha$=0°) or land to sea ($\alpha$=180°) at $M_g$=0.4, as depicted in Figs. 3 and 4, the three corresponding subplots in each figure ($M_g$=0.4: $\Delta\theta_1$=0, $\Delta\theta_1$=5, $\Delta\theta$=10) all belong to the transitional LSB regime. However, when the land is about to start cooling, only the advanced cooling subplot of Fig. 3 ($\alpha$=0°, $M_g$=0.4: $\Delta\theta_2$=−5) and the subplots of Fig. 4 ($\alpha$=180°, $M_g$=0.4: $\Delta\theta_2$=0, $\Delta\theta_1$=−5, $\Delta\theta_2$=−5) qualify as transitional LSBs. In a nutshell, such a regime only exists at weak synoptic forcings, i.e. for $M_g$<=0.4, and it is featured in both directions ($\alpha$=0° and 180°) but at different times throughout the day since it depends on the strength and sign of the thermal contrast $\Delta\theta(t)$.

*4.3 Land-driven thermal circulation*

A different regime arises when the synoptic wind blows from land to sea ($\alpha$=180°), the land-driven LSB, as identified in ABI23. This regime features the development of a shallow persistent SB only along this direction ($\alpha$=180°), but this is conditioned on the strength of the pressure forcing $M_g$ and $\Delta\theta(t)$. One remark here is that the transient dynamics feature this regime at $M_g$= 1.2 and 2 m/s only in the subplots corresponding to ($\alpha$=180°: $\Delta\theta$=10, $\Delta\theta_2$=5, $\Delta\theta_2$=0) of Fig. 4.

*4.4 Advected thermal circulation*

Advected LSBs display an almost unidirectional flow across the whole ABL domain; therefore, such flows do not feature any clear circulation cell, yet they might inherit some weak remnants of freshly eradicated LSBs. The flow in such regimes is nearly pressure driven, with almost no return air. This is clearly discerned along both directions, with the direction of the advected LSB dictated by $\alpha$. The subplots of Fig. 3 along ($\alpha=0°$, $M_g=1.2$: $\Delta\theta_1=0$, $\Delta\theta_1=5$, $\Delta\theta=10$, $\Delta\theta=-10$ and $\Delta\theta_2=-5$) and ($\alpha=0°$, $M_g=2$: for all $\Delta\theta$) demonstrate this regime i.e., "Advected-SL: from sea to land". Similarly, the subplots of Fig. 4 ($\alpha=180°$, $M_g=1.2$ and 2: $\Delta\theta_1=0$, $\Delta\theta_1=5$, $\Delta\theta_1=-5$, $\Delta\theta=-10$ and $\Delta\theta_2=-5$) show this regime i.e., "Advected-LS: from land to sea this time".

## 5 Thermal circulation structures: Evolving dynamics in time

Movies of the temperature and velocity field for both angles and selected geostrophic winds are available at the following links: they show the full 72 hours of simulations that include the warmup periods. Pseudocolor plot movies for the along shore, time averaged streamwise velocity and temperature ($\bar{U}_{y,t}$ and $\bar{\theta}_{y,t}$) through an X-Z slice of the analysis domain were generated for $M_g=0$ and for each $\alpha$ ($\alpha=0°$ and $\alpha=180°$) in increasing $M_g$ ($M_g=0.4$, 1.2 and 2 m/s). Overbar will henceforth denote along-shore averaging in Y and in time over 12.5 minutes prior to the time or $\Delta\theta(t)$ indicated in the plots. The corresponding color bar ranges are the same ones used in generating Figs. 3 and 4.

- Temperature and velocity field at $M_g = 0$ m/s

https://www.dropbox.com/s/fbgfrsl7bqdmpj6/Mg0_mov_mixed1p_U_T%20Mohammad%20Allouche.mov?dl=0

- Temperature field at $\alpha = 0°$:

https://www.dropbox.com/s/217xlcdri88j056/alpha0_mov_Mgs_1p_T%20Mohammad%20Allouche.mov?dl=0

- Streamwise velocity field at $\alpha = 0°$:

https://www.dropbox.com/s/hou20d6skbpp287/alpha0_mov_Mgs_1p_U%20Mohammad%20Allouche.mov?dl=0

- Temperature field at $\alpha = 180°$:

https://www.dropbox.com/s/3qnkbtk9takq4yn/alpha180_mov_Mgs_1p_T%20Mohammad%20Allouche.mov?dl=0

- Streamwise velocity field at $\alpha = 180°$:

https://www.dropbox.com/s/6n5uz17zo1wqqji/alpha180_mov_Mgs_1p_U%20Mohammad%20Allouche.mov?dl=0

It is recommended to watch the movies before or while reading the rest of this section.

## 5.1 Transient LSBs along $M_g=0$

The subplot ($M_g=0$, $\Delta\theta_1=5$) in Fig. 3 depicts a canonical SB that is well established over the whole analysis domain. Further increase in the thermal contrast would favor two simultaneous mechanisms. First, the canonical SB is expected to intensify as buoyancy is enhanced, and second, the convective thermals would start to develop over land. The interaction between these newly born thermals and the SB is noticed in the {$M_g=0$, $\Delta\theta=10$} subplot, but it is more discernable in the {$M_g=0$, $\Delta\theta_2=5$} subplot where the matured convective rolls infiltrate through the already established SB and modify its onshore penetration distance. After that time, the weakening of buoyancy over land at {$M_g=0$, $\Delta\theta_2=0$} causes the SB to reintensify rapidly and the thermals to lose structure, thus the SB's front pushes inland again. In the subplot {$M_g=0$, $\Delta\theta_1=-5$}, and although the temperature contrast sign is reversed here, the SB's thermal and kinetic inertias cause it to continue to expand up to the end of the domain, rendering the land stable, and gaining its canonical shape again. The tendency here is for the land to become more stable as it cools, and this is aided by the mildly stable background when $\Delta\theta$ first becomes negative because of the SB penetration.

When $\Delta\theta_1=-5$, the existing SB has not been damped yet because suppressing its kinetic energy solely by buoyancy destruction (attributed to stable ABL due to nocturnal cooling) is a slower process (than its generation) due to the stable stratification over land that reduces friction and heat exchange (especially that here the land is kept relatively smooth, so mechanical production of turbulent kinetic energy is weak). Further reducing the surface thermal contrast favors the formation of a more intensified internal stable boundary layer over land. Thus, a significant negative surface $\Delta\theta$ ($\Delta\theta=-10$) induces land breeze initiation only near the surface, slowly expanding vertically, remaining shallow and below the previously established SB from the daytime, as depicted in the subplot {$M_g=0$, $\Delta\theta=-10$}. Sustained colder land temperatures finally fully deplete the sea breeze, establishing a canonical land breeze later during the night, as shown in the subplot {$M_g=0$, $\Delta\theta_2=-5$}.

Completing the cycle in the subplot {$M_g$=0, $\Delta\theta_1$=0} in the top row, the land breeze prevails before the land becomes warmer than the sea. As the cycle restarts and the land warms {$M_g$=0, $\Delta\theta_1$=5}, the bulk circulation reverses more rapidly than during the day-to-night transition. The cold air remaining from the nighttime over land delays the onset of thermal convection. Thus, the intrinsic LSB's thermal and kinetic memory effects delay the land stability response, and the development of convection does not start until $\Delta\theta$=10, when the overlying LSB stable boundary layer is eradicated. Hence, one should be careful in analyzing LSBs as they are highly sensitive to the initial thermal and kinetic pervious states, especially under weak synoptic forcings background.

The top of Fig. 5 shows the correlation coefficient $R(t)$ between the two dimensional ($x$ and $z$) fields with the transient $U(x,z)$ simulated here, and its corresponding steady state "SS" scenario we simulated in our previous work. As could be anticipated, $R(t)$ increases with time, and one could anticipate that it should tend towards 1 as time approaches 6 h, the point where $\theta_L$=$\theta_{max}$ "SS". However, $R(t)$ reaches a local maximum of 0.85 earlier (when $\Delta\theta_1 \approx 6.75$ K), plateaus, and then starts decreasing. Here, $R(t)$ reaches a local minimum of 0.56 because of the SB interaction with the convective rolls (that are not seen in the SS cases), plateaus, and then starts increasing again as the thermals decay. $R(t)$ attains the global maximum of $R_{max}$=0.97 (almost one), around 7 hours following the SS time ($\tau_{lag} \approx$ 7 h), when $\Delta\theta_1 \approx -3.25$ K and the surface temperature contrast has already reversed. Afterwards, and as expected $R(t)$ starts to decorrelate during night times ($\Delta\theta(t) \ll 0$) where it reaches a global minimum of around −0.79 corresponding to a canonical LB ( mirror of a canonical SB).

The bottom of Fig. 5 shows the corresponding averaged enstrophy "a physical measure of the circulation strength" transient behavior $\varepsilon(t)$ relative to the SS value corresponding to $\theta_L$=$\theta_{max}$. At the corresponding time to $\theta_L$=$\theta_{max}$, the transient averaged enstrophy comprises only around 16% of the SS value, and then starts ramping up as $R(t)$ approaches $R_{max}$. Remarkably, it can be clearly seen that enstrophy reaches the global maximum ($\varepsilon_{max}$ is almost twice the SS value) exactly at the same time ($\tau_{lag}$) corresponding to $R$ =$R_{max}$, as indicated by the vertical magenta dotted line.

Fig. 6 shows three snapshots of the time averaged stream-wise $\langle U \rangle_{y,t}$ for the SS case of $\theta_L=\theta_{max}$ (top), the TR case "TR: transient" at the time corresponding to $\theta_L=\theta_{max}$ (middle), and TR case at the time corresponding to $R=R_{max}$ (bottom). The bottom TR case clearly reflects the main SS features (top), but it developed with a $\tau_{lag} \approx 7$ h relative to the SS time ($\Delta\theta_1 \approx -3.25$ K) with our prescribed time-varying land surface temperature (which is idealized and different from the real world where it would be controlled by the surface thermal inertia and energy budget).

*5.2  Transient LSBs with synoptic forcing along α=0°*

Here a sea-to-land pressure gradient is added, along with *Y-Z* inflow slices of the dynamics and flow structures of an infinite neutral flow over sea. The subplot $\{M_g=0.4, \Delta\theta_1=0\}$ in Fig. 3 demonstrates a transitional regime where traces of the night land breeze are still discerned especially near the surface. This is accompanied by an inherent stable boundary layer over land (away from the surface) and offshore (just near the shore where the LB prevails), and almost a neutral ABL over the sea (because of the advected inflows). As the land heats up, in the subplot ($M_g=0.4, \Delta\theta_1=5$), mixing is more favored over land and the unstable boundary layer near the surface deepens and weakens the stably stratified overlaying layer.

The transient behavior after this time is very similar to $M_g=0$ up to ($M_g=0.4, \Delta\theta=-10$), but the LSBs are more accelerated because of the directionality of the advected inflows along $\alpha=0°$. Further decrease of the thermal contrast in the subplot ($M_g=0.4, \Delta\theta_2=-5$) results in a transitional regime where the canonical LB loses its shape as result of the advected inflows that are trying to keep the sea neutral. Therefore, any canonical LB is short-lived along this direction. Also, canonical SBs require more heating over land because the sea tends to establish neutral stratification (because of inflows), and therefore the heat flux contrast across the whole ABL depth between land and sea is diminished compared to the $M_g=0$ transient behavior.

The transient dynamics for larger pressure forcings (subplots of $M_g=1.2$ and 2) result in almost advected LSBs (exceptions for $\Delta\theta_2=5$, $\Delta\theta_2=0$, $\Delta\theta_1=-5$ at $M_g=1.2$). LSBs weakly resist the pronounced increase of the pressure forcing here as the air over the sea remains neutral, and the flow over land goes from mildly to extremely unstable. However, these LSBs are boosted by buoyancy as their strength exceeds sometimes $M_g$. One remark here (for both $M_g$ =1.2 and 2) is that the canonical LB does not form any more; instead, an extremely persistent shallow micro-LB forms near the surface at night. This micro-LB is similar to a density current that could trigger turbulence intermittency at the finest scales over land and nearby sea (see Mahrt (2010) Allouche et al. (2022) for other intermittency triggering mechanisms). The LB's leading edge protrudes the furthest into the sea for the maximum buoyant destruction attained at $\Delta\theta_2=-5$, as it resists the strengthened incoming inflows.

The top panel in Fig. 7 (like Fig. 5), shows $R(t)$ and the averaged enstrophy $\varepsilon(t)$ for $\alpha=0°$ (here the correlation is relative to the steady cases with the same $M_g$ and $\alpha$). $R(t)$ increases gradually following the time corresponding to $\theta_L=\theta_{max}$ and eventually reaches $R_{max}=0.15$ for $M_g$ =0.4, $R_{max}=0.81$ for $M_g$ =1.2, and $R_{max}=0.85$ for $M_g$ =2. As expected, $R(t)$ attains higher maximum values as $M_g$ increases (transient thermal dynamics become less important). Morevoer, at weaker $M_g$, the likelihood of recovering the corresponding steady state LSB is limited since memory effects play a critical role in modulating the resultant dynamics. All cases, however, have a comparable time lag to peak correlation of about $\tau_{lag}$=7-hour relative to the SS circulation to $\theta_L=\theta_{max}$. Almost, the same $\tau_{lag}$ reported for $M_g$ =0 (subsection 5.1) qualifies here as well for $\alpha=0°$ at all $M_g$. We also note here that the averaged enstrophy $\varepsilon(t)$ trends (for all $M_g$) peak almost at the time corresponding to such a characteristic lag ($\tau_{lag}$). Beyond this lag, both $R(t)$ and $\varepsilon(t)$ drop significantly.

## 5.3 Transient LSBs along α=180°

Here, a land-to-sea pressure gradient is imposed, along with Y-Z inflow slices of the dynamics and thermal structures of diurnal flow over an infinite land. The subplot {$M_g$ =0.4, $\Delta\theta_1$=0} in Fig. 4 demonstrates a transitional regime where residuals of the night land breeze are still observed (the LB here is much stronger than the LB corresponding to $\alpha=0°$ as it is aided by the inflows). One important remark is that the leading edge of this land breeze prevails further downstream over the sea compared to $\alpha=0°$.

As the land heats up, accompanied by the advection of strengthened unstable structures in the subplot ($M_g$ =0.4, $\Delta\theta_1$=5), mixing is increased over land and the unstable boundary layer near the surface deepens more rapidly. These simultaneous mechanisms facilitate the eradication of the mildly stably stratified overlaying layer. With further increase in the thermal contrast in the subplot ($M_g$ =0.4, $\Delta\theta$=10), SB initiation starts here earlier compared to $\alpha$=0°, but the advection of thermal convective rolls from land to sea hinders the SB front inland penetration where it stalls near the shore at $X$=43 km. The SB hardly penetrates in the subplot ($M_g$ =0.4, $\Delta\theta_2$=5), as it is resisted by the advected structures. In the subsequent subplot ($M_g$ =0.4, $\Delta\theta_2$=0), the SB continues to strengthen as buoyancy here is still being boosted inland ($\Delta\theta$ >=0) and diminished over the sea patch because the advected inflows render the sea more stable with respect to time, and therefore this makes the SB more persistent. In the subplot ($M_g$ =0.4, $\Delta\theta_1$=−5), the LSB continues to transition as the potential LB formation is dynamically and thermally aided near the surface by the stably advected inflows. In the subplot ($M_g$ =0.4, $\Delta\theta$=−10), a canonical LB forms, which then begins to fade in the subsequent subplot ($M_g$ =0.4, $\Delta\theta_2$=−5) as stability over land is weakened. The advected inflows at this stage easily destroy this canonical LB and replaces it with a transition regime LSB in subplot ($M_g$ =0.4, $\Delta\theta_1$=0).

The transient dynamics for larger pressure forcings (subplots of $M_g$ =1.2 and 2) result only in land-driven SBs for $\Delta\theta$=10, $\Delta\theta_2$=5, and $\Delta\theta_2$=0 for both $M_g$'s and advected LSBs for all other $\Delta\theta$'s (canonical LSBs do not form anymore). In a nutshell (along $\alpha$=180°), as the land heats during the day, amplified advection of unstable inflows ensues and renders the land more unstable and the sea more stable. Therefore, SB initiation is favored to start earlier (compared to $\alpha$=0°), and as the pressure forcing increases, it persists longer (as this correlates with stronger stabilities over the sea) but its front is strongly impeded inland (because of the strengthened opposing inflows). During nighttime, as the land starts cooling, aided by the stably advected inflows, the sea inherits a stably stratified thermal memory that inhibits its regime shift until much later into the night, and thus remains almost neutral at that stage. Therefore, LB formation is not favored (especially in its canonical shape).

The bottom panel in Fig. 7, shows $R(t)$ and the averaged enstrophy $\varepsilon(t)$ for $\alpha$=180°. For $M_g$ =0.4, $R(t)$ increases following the time corresponding to $\theta_L=\theta_{max}$ and reaches $R_{max}$=0.78 >> $R_{max}$=0.15 for $\alpha$=0°. For $M_g$ =1.2 and 2 respectively, $R(t)$ almost plateaus during this time and reaches $R_{max}$=0.41 << $R_{max}$=0.81 for $\alpha$=0°, and $R_{max}$=0.68 < $R_{max}$=0.85 for $\alpha$=0°. $R_{max}$

values here change in a non-monotonic sense as $M_g$ increases. An approximate characteristic lag of $\tau_{lag}$=5-hour relative to the SS time corresponding to $\theta_L=\theta_{max}$ is found here. The fact that ($\tau_{lag}$=5-hour here for $\alpha$=180°) < ($\tau_{lag}$=7-hour for $\alpha$=0° and for $M_g$=0) can be attributed to the nature of these transient inflows, which causes land-driven SBs to initiate earlier as described in the transient dynamics above, and thus faster recovery of the corresponding SS LSB if achieved. Remarkably here for $\alpha$=180° and unlike $\alpha$=0°, beyond the identified lag (that is for $t > \tau_{lag}$=5-hour), the drop in $R(t)$ is accompanied by a sharp increase in the averaged enstrophy $\varepsilon(t)$ for all $M_g$.

## 6 Surface and shore exchanges

### 6.1 Surface exchanges

We plot in Fig. 8 the diurnal profiles of the surface heat flux ($\overline{w'\theta'}$) for both directions ($\alpha$=0° and $\alpha$=180°) over the two analysis patches, sea and land. For $M_g$=0, and as described in the previous subsection 5.1, $\overline{w'\theta'}_{Sea}$ is positive (unstable) at $t$=0 before the land starts heating up because of the prevailing cool nighttime land breeze. As we progress in time, $\overline{w'\theta'}_{Sea}$ drops to almost zero at $t$=6 h ($\Delta\theta=\Delta\theta_{max}$), i.e., a neutral sea. Note that the sea remains neutral at such thermal contrast and for most of the daytime, while in the steady state cases it is slightly stable. This is attributed to the memory effects of the persistent thermal boundary layer associated with the previously established near-surface land breeze. At $t$=18 h ($\Delta\theta=-\Delta\theta_{max}$), a canonical land breeze develops, and the sea returns gradually to an unstable condition at the end of the cycle at $t$=24 h ($\Delta\theta$=0). Therefore, $\overline{w'\theta'}_{Sea}$ reveals a strong hysteresis for the $M_g$=0.

Along $\alpha$=0° and as $M_g$ increases ($M_g$=0.4 ms$^{-1}$ subplot), this hysteretic response is weakened especially when $\Delta\theta>0$, and the observed $\overline{w'\theta'}_{Sea}$ unstable maximum value reduces by almost one order of magnitude (relative to $M_g$=0 ms$^{-1}$) because of the neutral sea advected inflows. For $M_g$=2, the hysteresis is eliminated when $\Delta\theta>0$, but persists marginally when $\Delta\theta<0$ because of the land breeze transient penetration over the sea. An important note here is that, if we average the sea surface heat flux by excluding dynamically (with respect to time) the sea sub patch where this land breeze prevails, the hysteresis response would be fully eliminated when ($\Delta\theta<0$, not shown here). This implies that the hysteresis is directly linked to the breeze dynamics near the shore. The friction velocity, $\bar{u}_{*Sea}$, experiences similar trends, but displays weaker hysteresis compared to $\overline{w'\theta'}_{Sea}$, and such type of response is completely eliminated for $M_g$=2 (not shown here).

In the same Fig. 8, over land with $M_g$=0, $\overline{w'\theta'}_{Land}$ is positive (unstable) when $\Delta\theta$>0 and negative (stable) almost for every point when $\Delta\theta$<0. The range of variability here for $\overline{w'\theta'}_{Land}$ is almost four times greater than $\overline{w'\theta'}_{Sea}$, and a much weaker hysteresis is observed over land. The relatively higher fluxes over land imply that land is the primarily driver of these LSBs, whereas the sea is adaptively responding to the former's transient dynamics. Note that the reported $\overline{w'\theta'}_{Land} \approx 0.06$ K m/s at $t$=6 h ($\Delta\theta$=$\Delta\theta_{max}$) is almost the same as the steady state value we observed in ABI23. As $M_g$ increases along $\alpha$=0° ($M_g$=0.4 subplot), the hysteresis is mildly strengthened especially when ($\Delta\theta$>0) because both buoyancy and shear are co-modulating the transient LSBs. For $M_g$=2, such type of response is almost fully eliminated. For $\alpha$=0° and $M_g$>0, the reported heat fluxes $\overline{w'\theta'}_{Land} \approx 0.06$ K m/s at $t$=6 h ($\Delta\theta$=$\Delta\theta_{max}$) are almost half the steady state reported values because $\Delta\theta$ is not \maintained at its maximum $\Delta\theta_{max}$. The friction velocity, $\bar{u}_{*Land}$ reveals stronger hysteresis response compared to $\overline{w'\theta'}_{Land}$, and these responses are never eliminated even for $M_g$=2 (not shown here).

Now along $\alpha$=180°, as $M_g$ increases ($M_g$=0.4 subplot), $\overline{w'\theta'}_{Sea}$ shows a strengthened hysteretic response compared to $\alpha$=0°, and $\overline{w'\theta'}_{Sea} \approx -1.5 \times 10^{-3}$ K m/s reaches a new minimum attributed to the diurnal land inflows, which augments the negative heat flux over the sea (stable, refer to the physical analysis in subsection 5.3). For $M_g$=2, such response is strengthened further because of the intensified diurnal land inflow planes where $\overline{w'\theta'}_{Sea} \approx -5 \times 10^{-3}$ more than triples relative to $M_g$=0.4. For $M_g$=0.4 and $M_g$=2, the reported heat fluxes $\overline{w'\theta'}_{Sea} \approx -0.3 \times 10^{-3}$ K m/s and $\overline{w'\theta'}_{Sea} \approx -1.4 \times 10^{-3}$ K m/s at $t$=6 h ($\Delta\theta$=$\Delta\theta_{max}$) are around 5% and 12% the steady state values in ABI23, again because $\Delta\theta$ is not maintained at its maximum $\Delta\theta_{max}$. In addition, the friction velocity, $\bar{u}_{*Sea}$ experience a similar but weaker hysteresis (not shown here). Over the land patch and as $M_g$ increases here along $\alpha$=180° ($M_g$=0.4 subplot), $\overline{w'\theta'}_{Land}$ shows a weakened hysteresis response compared to $\alpha$=0° because of the diurnal land inflows along this direction especially when $\Delta\theta$>0. For $M_g$=2, the hysteresis response is almost fully eliminated as with the corresponding $\alpha=0°$ case. For $\alpha=180°$ and $M_g>0$, the reported heat fluxes $\overline{w'\theta'}_{Land} \approx$ 0.04 K m/s at $t=6$ h ($\Delta\theta=\Delta\theta_{max}$) are almost one third the steady state values. Also, the friction velocity, $\bar{u}_{*Land}$ shows stronger hysteresis responses compared to $\overline{w'\theta'}_{Land}$ (not shown here).

In a nutshell, the land surface heat flux hysteresis response is weakened as $M_g$ increases, and is almost eliminated for $M_g=2$. However, the sea surface heat flux hysteresis is strengthened for $M_g=2$ ($\alpha=180°$), and significantly weakened but never eliminated for $\alpha=0°$ because of the prevailing near surface land breeze here. Hence, a sea surface heat flux hysteresis is always exhibited anytime the diurnal LSBs are passing through any of the canonical reference states LSB types (canonical SB, canonical LB, land-driven SB). Note that purely diurnal advected LSBs might still show a very weak $\overline{w'\theta'}_{Sea}$ hysteretic response when micro LBs or SBs prevail. This is discerned for $M_g=2$ along $\alpha=0°$ when the shallow near surface LBs persist, or as could be anticipated for $M_g>2$ along $\alpha=180°$ since the shallow land-driven SBs would still form (not simulated here).

## 6.2 Mean flow and turbulence at the shore

$Q_{shore}$ is defined here as the net volumetric flux across a unit along-shore width, $\|y_n\|=1$ m (added merely for convenience in interpreting units), at the shore interface $X=40$ km. $Q_{sc}$ represents a normalized version of $Q_{shore}$, and both are defined below:

$$Q_{shore} = \frac{\|y_n\|}{L_y} \int_0^{L_y} \int_{z_0}^{L_z} U_{shore} dz\, dy, \qquad (4)$$

$$Q_{sc} = \frac{Q_{shore}}{M_g L_z \|y_n\|} = \frac{1}{L_z L_y M_g} \int_0^{L_y} \int_{z_0}^{L_z} U_{shore} dz\, dy. \qquad (5)$$

Fig. 9 shows the diurnal variation $Q_{sc}$ for each $\alpha$ with increasing $M_g$, with respect to the surface forcing $\Delta\theta(t)$ and turbulent kinetic energy at the shore ($e_{shore}$, averaged in the same way as $Q_{shore}$). We do not plot the case with $M_g = 0$ since the shore flux in these is nearly zero. Along $\alpha=0°$ and as $M_g$ increases, the $\{Q_{sc}, \Delta\theta(t)\}$ and $\{Q_{sc}, e_{shore}\}$ hysteretic responses are weakened and then almost eliminated for $M_g=2$. One can even observe one case of quasi-equilibrium for $\alpha=0°$ and $M_g=2$. However, along $\alpha=180°$ and as $M_g$ increases, the $\{Q_{sc}, \Delta\theta(t)\}$ and $\{Q_{sc}, e_{shore}\}$ responses are strengthened and more hysteretic, and we do not discern any case of quasi-equilibrium for the simulated cases.

For both directions, the mean flow is out of equilibrium with turbulence anytime the sea surface heat flux shows a significant hysteretic behavior as depicted in Fig. 8. The fact that the land patch exhibits stronger surface heat fluxes than the sea and much weaker hysteretic responses as $M_g$ increases does not guarantee attaining turbulence equilibrium with the mean; therefore, the final turbulence-mean flow equilibrium response is quite sensitive to the sea surface heat flux behavior as well.

Note that the reported $Q_{sc}$ values along $\alpha=0°$ at $t=6$ h ($\Delta\theta=\Delta\theta_{max}$) $\{Q_{sc}=1.29$ for $M_g=0.4$, $Q_{sc}=1.08$ for $M_g=1.2$, $Q_{sc}=1.05$ for $M_g=2\}$ are almost the same as their corresponding steady state reported values. Along $\alpha=180°$, the reported $Q_{sc}$ values at $t=6$ h ($\Delta\theta=\Delta\theta_{max}$) $\{Q_{sc}=-0.72$ for $M_g=0.4$, $Q_{sc}=-0.86$ for $M_g=1.2$, $Q_{sc}=-0.91$ for $M_g=2\}$ are slightly different than their corresponding steady state values $\{Q_{sc}=-0.69$ for $M_g=0.4$, $Q_{sc}=-0.91$ for $M_g=1.2$, $Q_{sc}=-0.95$ for $M_g=2\}$ because of the inherent memory effects and the nature of the continuously evolving inflows fed in the main unsteady simulation. For $M_g>0.4$, both the steady state and unsteady simulations result in land driven LSBs at $t=6$ h, but more mass flux occurs in the steady state simulations because of the fact that $\Delta\theta$ is maintained at its maximum $\Delta\theta_{max}$, which results in shallower land-driven LSBs.

## 7 Macro and micro LSBs categorization

Visual inspection and previous analysis of the thermal circulations large scale flow structures in sections 4 and 5 reveal four types of LSBs: (i) canonical LSBs of two flavors, Canonical-CCW SB (CCW: counter-clockwise LSB) and Canonical-CW LB (CW: clockwise LSB); (ii) a Transitional LSB, (iii) a Land-driven SB, and (iv) advected LSBs of two flavors, Advected-SL LSBs (SL: sea to land) or Advected-LS LSBs (LS: land to sea). Therefore, we consider the vertical profile of the shore-normal velocity $\overline{U}(z)$ at the shoreline, $\overline{U}(z)_{shore}$, with the aid of the *k*-means clustering algorithm (Lloyd, 1982), to delineate these circulations without any human visual inspection. This autonomous algorithm is provided with $\overline{U}(z)_{shore}$ as an input (only for the last two cycle to have more training data), and the number of the desired clusters (= 6 as depicted) is specified as well.

The top panel of Fig. 10 shows the macro-classification results that are overlaid on the land surface temperature profile to easily track the evolution of the identified regimes. As could be inferred from the movies and section 5, the clustered data here (as learnt from the vertical profiles of $\overline{U}(z)_{shore}$) reflect their corresponding regimes most of the time, except when the LSB is about to shift regime and the classification is more uncertain.

The top panel of Fig. 10 depicts the large-scale evolution of the clustered LSBs in the entire ABL, but that does not necessarily reflect what is happening very near the surface. Therefore, in the bottom panel of Fig. 10, we consider the streamwise profile of the shore-normal velocity $\overline{U}(x) - z_1$ at the first grid level above the surface ($z_1$), also with the aid of *k*-means clustering algorithm to produce a micro-classification of the regimes. The number of the desired clusters is here set to 4 here because this corresponds to the observed air dynamics over the two patches as will be explained shortly. The exponent sign symbol here represents the air direction over each patch, and the number of signs denotes the relative strength in air dynamics: (i) $S^+L^+$: air is fully advected from sea to land through all $L_x$, (ii) $S^+L^{--}$: air over land tends to penetrate the sea patch, i.e., the prevailing nighttime shallow land breeze for the case of *α*=0°, $M_g$>0.4 (clearly seen in the movies), (iii) $S^{++}L^-$: air over sea tends to penetrate the land patch, i.e., the land-driven SB for the case of *α*=180°, mainly at $M_g$>0.4 (clearly seen in the movies), (iv) $S^-L^-$ : air is fully advected from land to sea through all $L_x$. The near surface air dynamics are fairly captured with such a $\overline{U}(x) - z_1$ micro-classification. One can note a correspondence of the bulk classification in the top panel of Fig. 10 with the surface micro classification in the bottom panel, but the timing of the transition and some details are not identical. This implies that surface measurements do not fully characterize the LSB regime.

## 8   Conclusion and implications

LES modeling of the unsteady LSBs, in the presence of constant synoptic pressure forcing along the symmetric directions *α*=0° and *α*=180°, reveals four different regimes: (i) canonical LSB, (ii) transitional LSB, (iii) land-driven SB, (iv) advected LSB. The first two types (i) and (ii) emerge when the synoptic forcing is weak ($M_g$=0 and 0.4) for both directions, and the LSB is shown to be transitional anytime it is shifting between the two canonical reference states "SB and LB". When the synoptic forcing strengthens along $\alpha=0°$, both canonical SB and advected LSB regimes are present at different times for $M_g$ =1.2, while only purely advected LSBs are found for $M_g$ =2. However, along $\alpha=180°$, land-driven SB and advected LSB regimes persist even up to the highest simulated case $M_g = 2$. In addition to the different interactions with the breeze circulations of the synoptic wind blowing from either direction, an important contributor to the difference for these transient cases is the asymmetry of the inherent inflows: for $\alpha=0°$ they consist of streamwise rolls with neutral stability, while and $\alpha=180°$, they feature stably stratified turbulence during nighttime and thermal convection during the unstable daytime. In addition, these inflows interact with an evolving diurnal thermal circulation that is modulated by its thermal and kinetic inertias, found to be influencial under weak synoptic forcings ($M_g$ =0 and 0.4) for both directions ($\alpha=0°$ and $\alpha=180°$).

Focusing on the transient dynamics of the cases with stronger pressure forcings ($M_g$ =1.2 and 2), along $\alpha=180°$, land-driven SBs initiation can start earlier in time (compared to $\alpha=0°$) because of the advected unstable inflows, but they fade away earlier too. In addition, land-driven SBs persists longer as $M_g$ increases along $\alpha=180°$ (because the strengthened land advected unstable inflows correlate with stronger negative heat fluxes over sea), whereas canonical SBs shift to an advected LSB reference type along $\alpha=0°$ (because the strengthened sea advected neutral inflows correlate with stronger positive heat fluxes over land). As the land starts cooling, along $\alpha=180°$, the sea still inherits a stable thermal memory that strengthens with time (because of the stably advected inflows), instead of weakening, and inhibits canonical LB formation. During the same time but along $\alpha=0°$, a micro (near surface) land breeze forms and prevails throughout the negative thermal contrast and beyond.

These dynamics often result in non-equilibrium flows over the two patches that manifest as a hysteresis in the response of the turbulence to the mean flow, as well as the response of the mean flow to the driving temperature contrast. This confirm that the mean flow has a significant memory that induces this hysteresis relative to the external forcing, and that the mean flow variability is too rapid for turbulence to equilibrate. Hysteresis is thus noted when the turbulent land surface fluxes over the diurnal cycle are plotted versus the temperature contrast, but this hysteresis is noticeably undermined when $M_g$ increases for both directions. On the other hand, the sea surface heat flux hysteresis relative to the temperature contrast is strengthened for $M_g$=2 at $\alpha$=180°, but substantially weakened yet never eliminated for $\alpha$=0° because of the persistent micro land breeze. The sea surface heat flux hysteresis persists anytime the diurnal LSBs pass through any of the canonical reference state LSB types (canonical SB, canonical LB, and land-driven SB). Advected LSBs are found to eliminate $\overline{w'\theta'}_{Sea}$ hysteresis, but any potential formation of either micro-SB or micro-LB during the day perturbs the response and pushes it back to a mild hysteretic type.

Similarly, the diurnal variation of the normalized net shore volumetric flux shows a hysteretic response for $\alpha$=180° as $M_g$ increases, but not for $\alpha$=0°. It is noticed that as $M_g$ increases (especially for $\alpha$=0°), i.e., the flow tends to be pressure gradient dominated, the hysteretic response of either surface heat fluxes ($\overline{w'\theta'}$, $\Delta\theta(t)$), or shoreline mean-flow ($Q_{sc}$, $\Delta\theta(t)$) and ($Q_{sc}$, $e_{shore}$) is undermined or almost eliminated. In fact, we observe only one case of equilibrium that corresponds to ($\alpha$=0° and $M_g$=2) in the simulated cases. The mean flow is found to be out of equilibrium with turbulence anytime the sea surface heat flux response shows a substantial hysteretic response with the external forcing despite the fact that the land patch converges to equilibrium faster as $M_g$ increases. For the simulated cases along $\alpha$=180°, we observe a strengthened hysteric response ($\overline{w'\theta'}_{Sea}$, $\Delta\theta(t)$) as $M_g$ increases because of the persistent land-driven SB, which in turn pushes the shoreline mean-flow response either ($Q_{sc}$, $\Delta\theta(t)$) or ($Q_{sc}$, $e_{shore}$) into non-equilibrium.

Finally, the vertical profile of the shore-normal velocity $\overline{U}(z)$ can be used with $k$-means clustering to delineate the large-scale evolution of the clustered LSBs in the entire ABL (macro categorization). Similarly, the stream-wise profile of the shore-normal velocity $\overline{U}(x) - z_1$ at the first grid level above the surface is used to demarcate the near-surface evolution of the LSBs (micro categorization). Both methods are tested, and they are found to complement each other in offering a descriptive picture of the resultant transient LSBs especially when micro-SBs or micro-LBs form with an advected LSB background regime.

This analysis offers insights into the dynamics and non-equilibrium of these LSBs on either side of the shore. The impacts on airflow, surface fluxes, cross-shore transport and multiple flow characteristics are significant and cannot be ignored in models that do not resolve this coastal sub-grid scale surface heterogeneity adequately. The potential implications of such a behavior are non-local because of the compounded and nonlinear interactions between rainfall, temperature, radiative effects, wind, humidity, and cloud formation during the time it takes the transient surface information to be communicated to the atmosphere. Incorporating these dynamics in such models (coarse weather and climate models) is thus an essential task, but as this study shows a challenging one due to the many embedded time-varying inputs, implicit memory effects and unresolved processes.

**Acknowledgments**

M.A. and E.B.Z. are supported by the Cooperative Institute for Modeling the Earth System at Princeton University under Award NA18OAR4320123 from the National Oceanic and Atmospheric Administration, and by the US National Science Foundation under award number AGS 2128345. The LES are conducted on the Cheyenne supercomputer of the National Center for Atmospheric Research (doi:10.5065/D6RX99HX) under projects UPRI0007 and UPRI0021. The authors declare no conflict of interest.

**Data availability**

The dataset of all the LES simulations are publicly available at https://doi.org/10.5281/zenodo.10433810 (Allouche et al. [2023]).

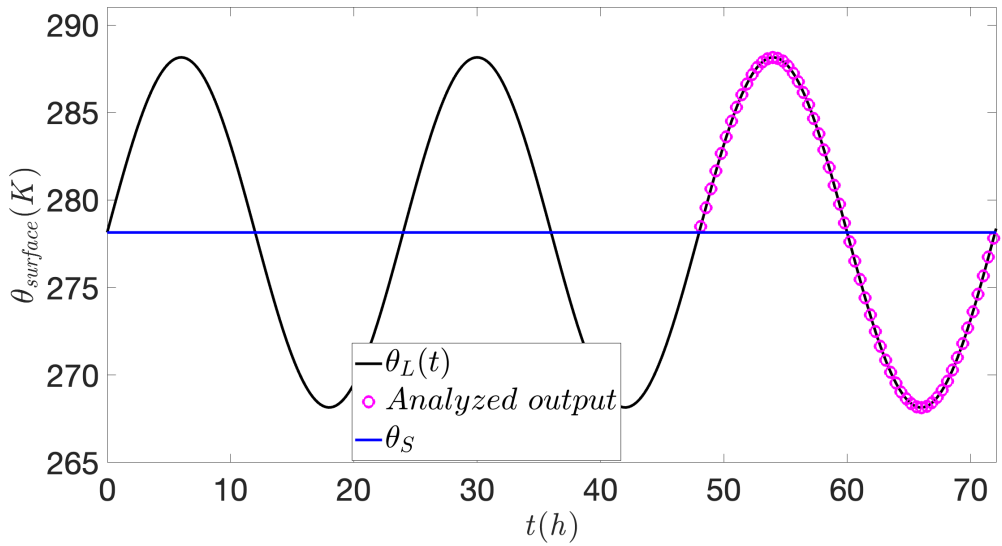
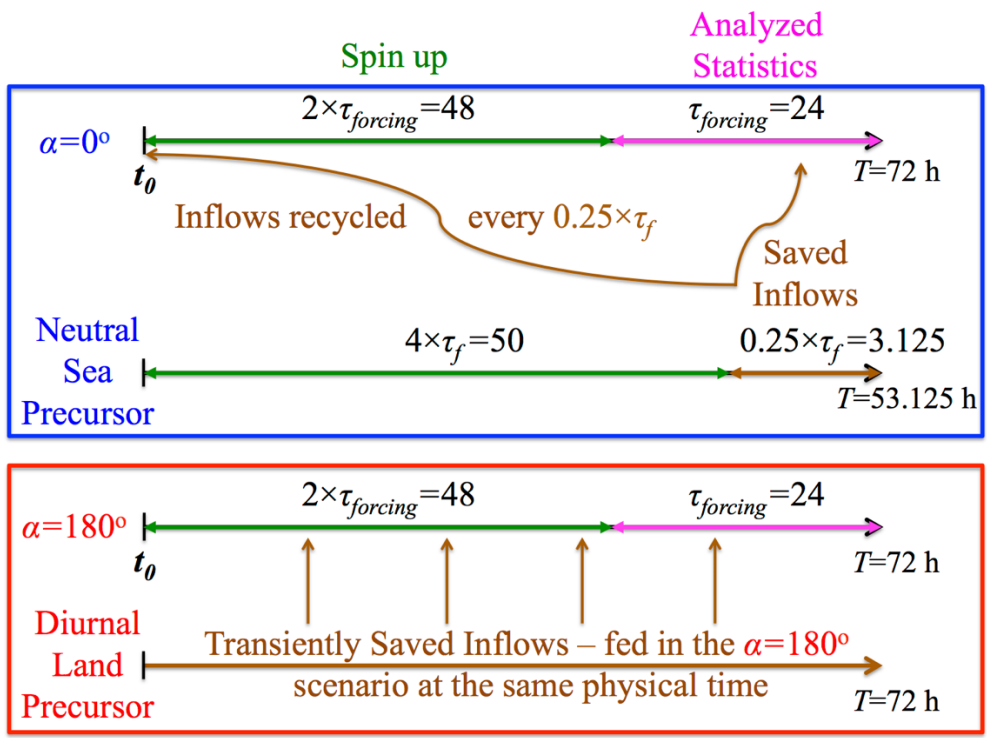

Fig. 1 The simulated diurnal surface temperature profile over the two patches. "Analyzed results" correspond to the third 24 hours period (top panel). Schematic of spin-up, inflows, and analyses periods for the $\alpha=0°$ and $\alpha=180°$ simulations (bottom panel)

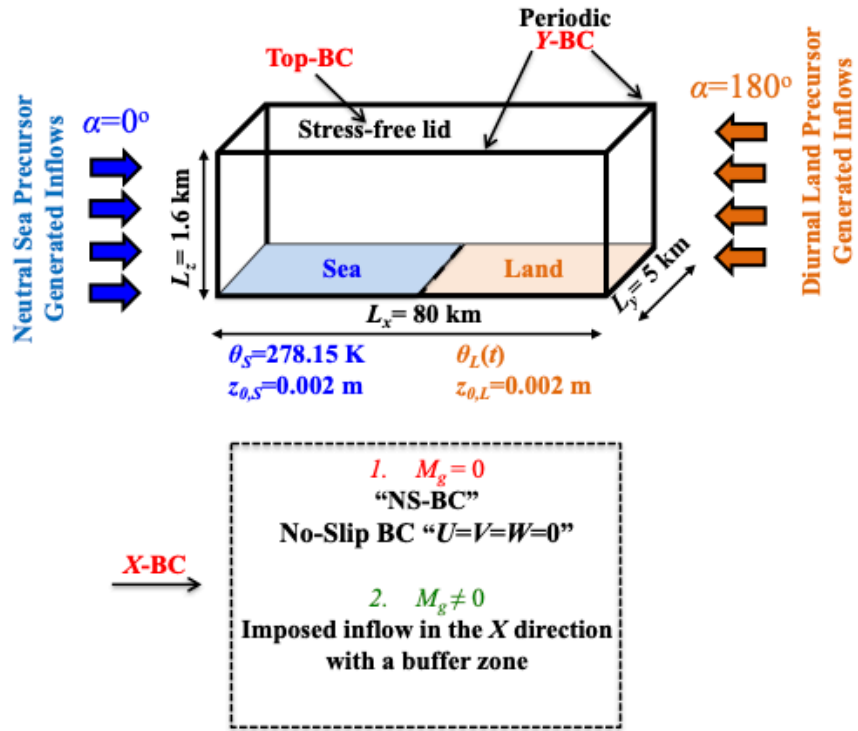

Fig. 2 Three dimensional schematic diagram of the sea-land heterogeneous simulation, BCs are fully detailed in ABI23

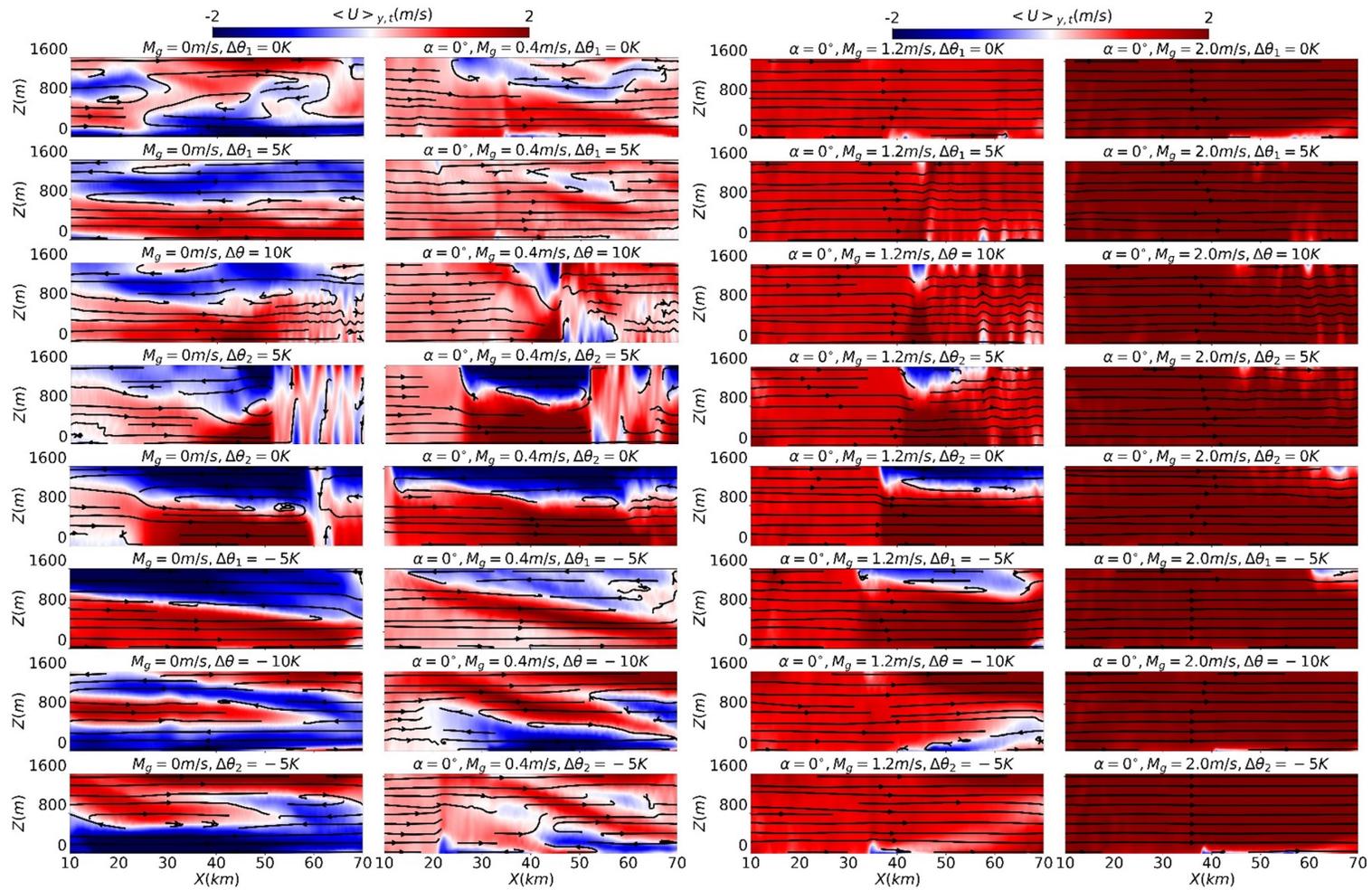

Fig. 3 Pseudocolor plots of along shore, time averaged stream-wise velocity ⟨$U$⟩$_{y,t}$ through an *X-Z* slice of the analysis domain.

The panels here correspond to $\alpha=0°$

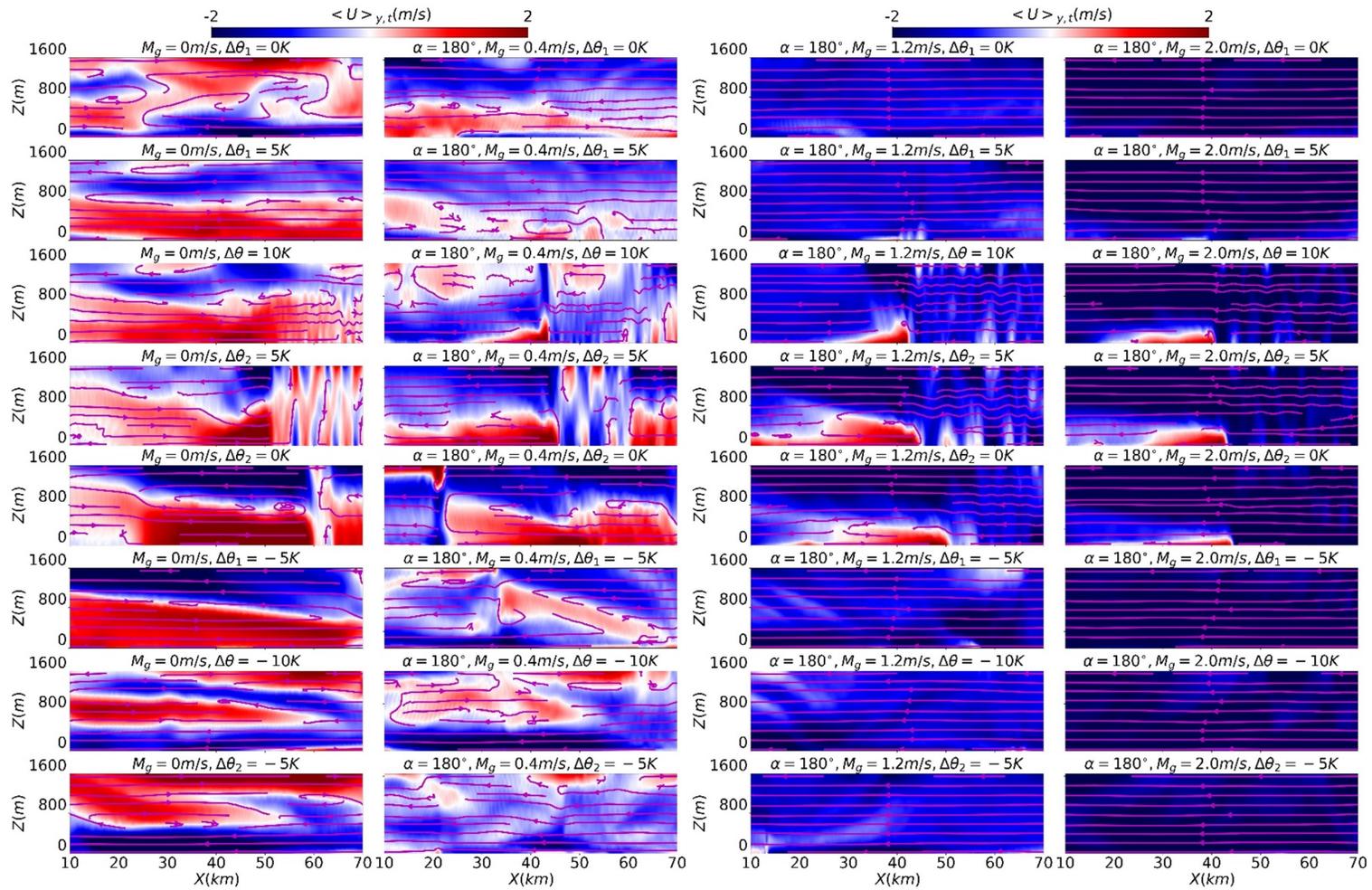

Fig. 4 Pseudocolor plots of along shore, time averaged stream-wise velocity $\langle U \rangle_{y,t}$ through an *X-Z* slice of the analysis domain. The panels here correspond to $\alpha=180°$

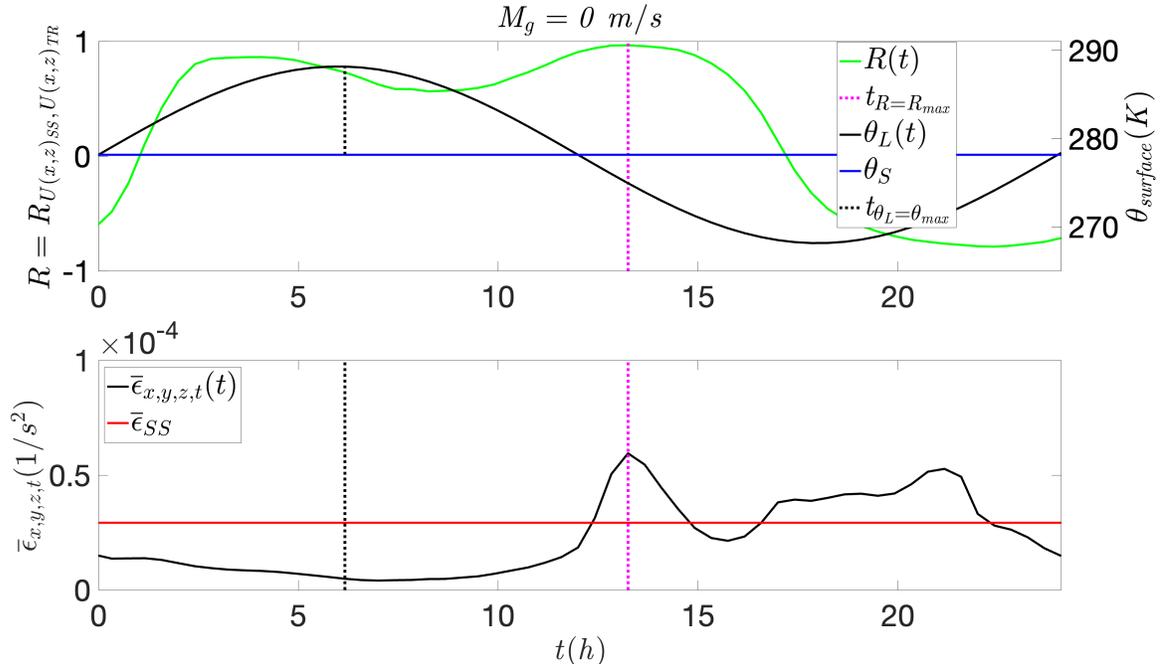

Fig. 5 The correlation coefficient transient behavior of ⟨U⟩$_{x,z}$ "subscripts: SS refers to one steady state snapshot when $\theta_L=\theta_{max}$, and TR refers to the transient snapshots of the unsteady simulation" for $M_g$=0 (top panel). The corresponding averaged enstrophy is shown here (bottom panel)

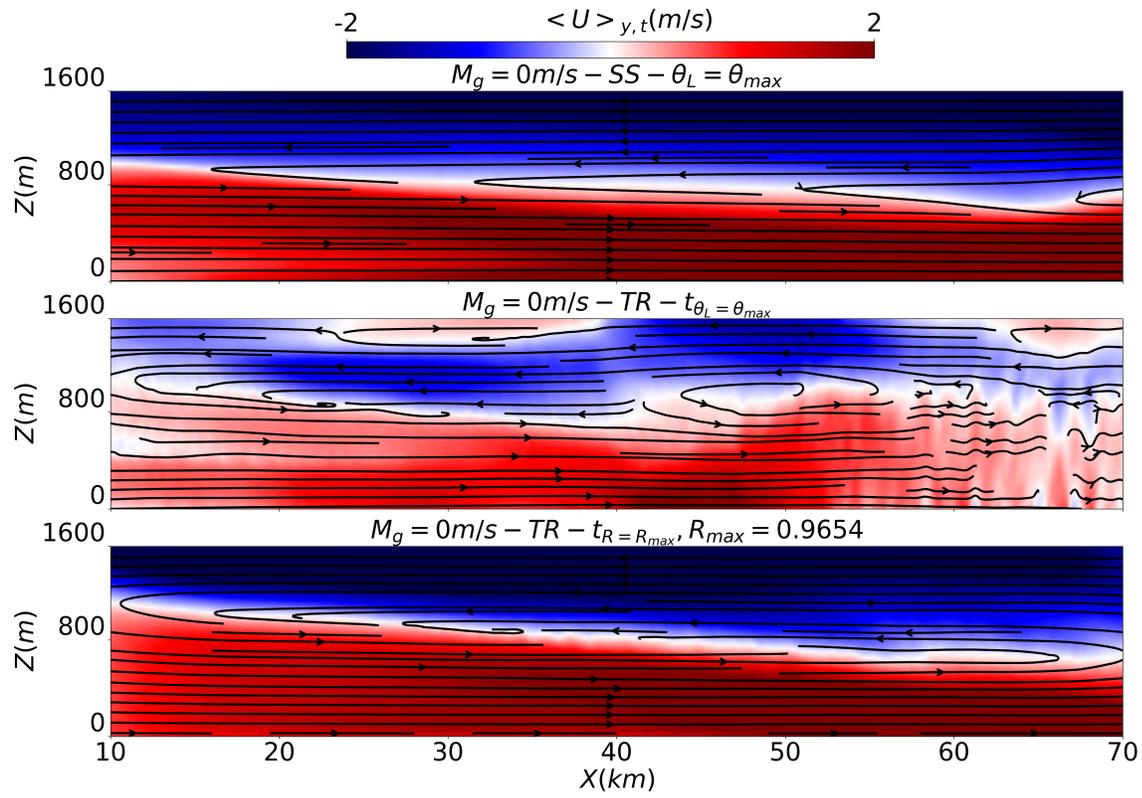

Fig. 6 The corresponding snapshots of ⟨U⟩$_{x,z}$ "SS: $\theta_L=\theta_{max}$" (top), "TR: $t_{\theta_L=\theta\,max}$" (middle), and "TR: $t_{R=Rmax}$" (bottom) for $M_g=0$

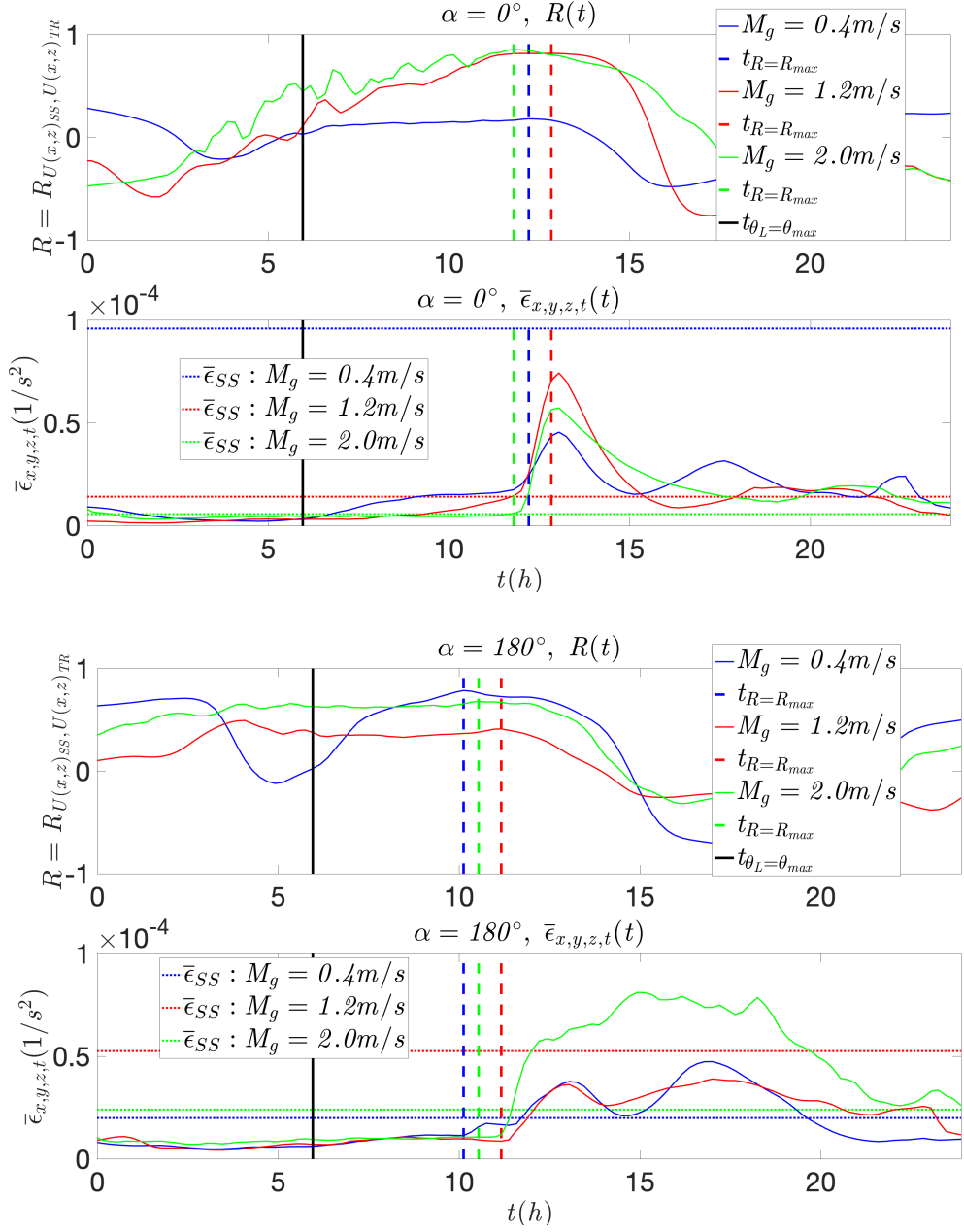

Fig. 7 The correlation coefficient transient behavior of $\langle U \rangle_{x,z}$ and the corresponding averaged enstrophy for $\alpha=0°$ (top panel). Similarly, for $\alpha=180°$ (bottom panel)

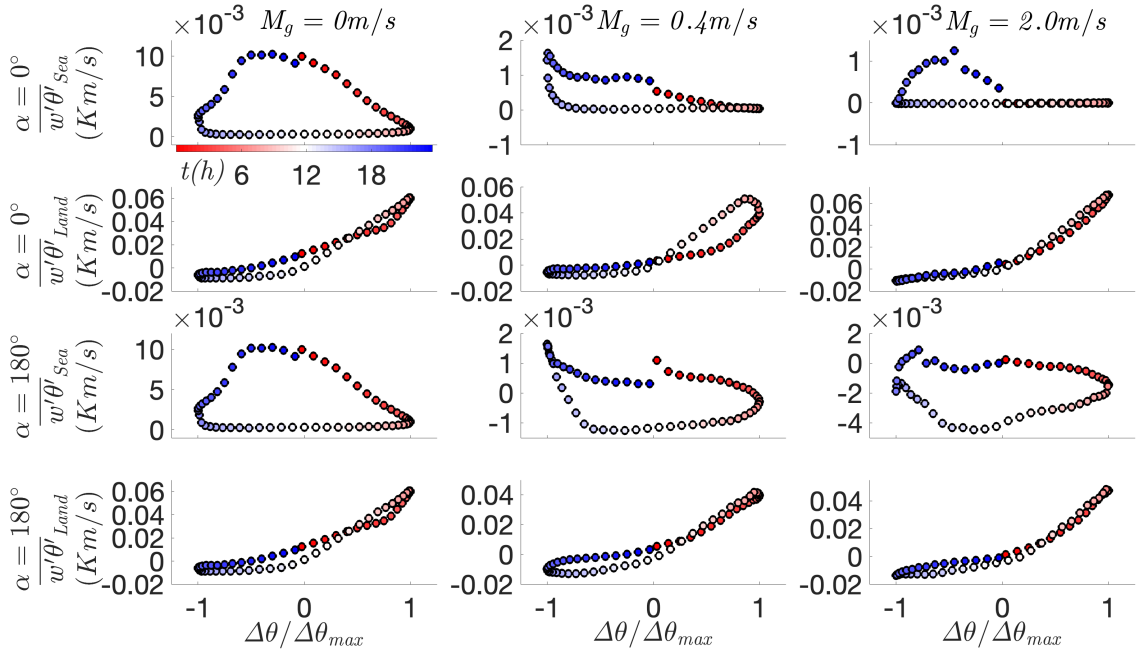

Fig. 8 The diurnal behavior of surface heat fluxes over the sea and land patches respectively with increasing $M_g$ along $α=0°$ (top two rows). Similarly, for $α=180°$ (bottom two rows)

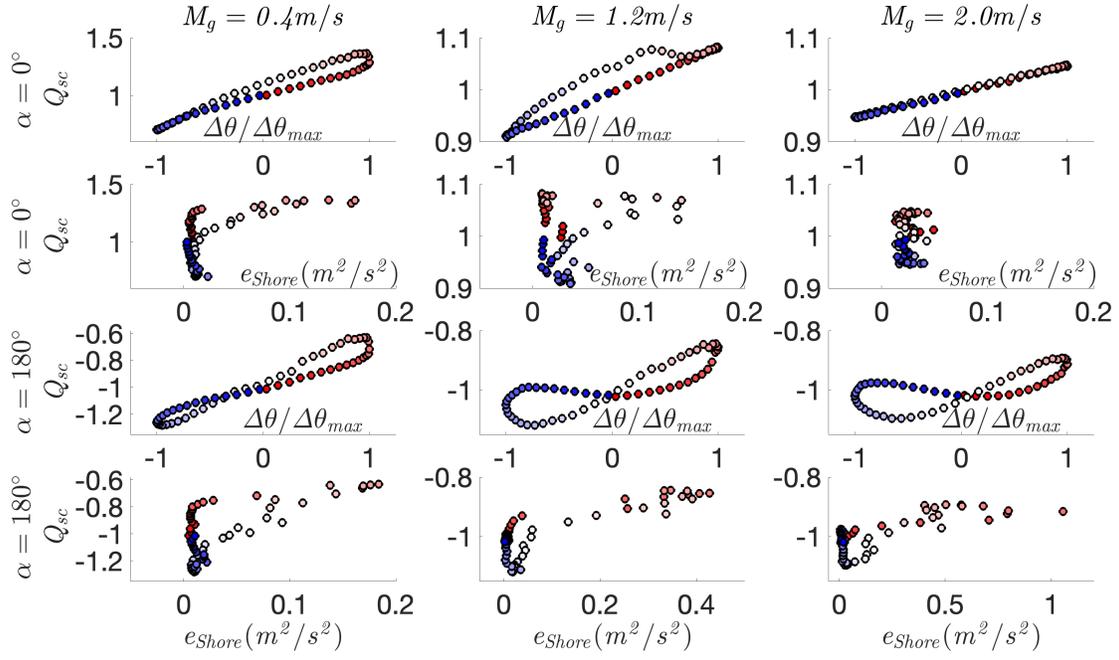

Fig. 9 The diurnal behavior of the normalized net shore volumetric flux ($Q_{sc}$) relative to the imposed forcing $\Delta\theta$ and TKE, $e_{Shore}$, with increasing $M_g$ along $\alpha=0°$ (top two rows). Similarly, for $\alpha=180°$ (bottom two rows).

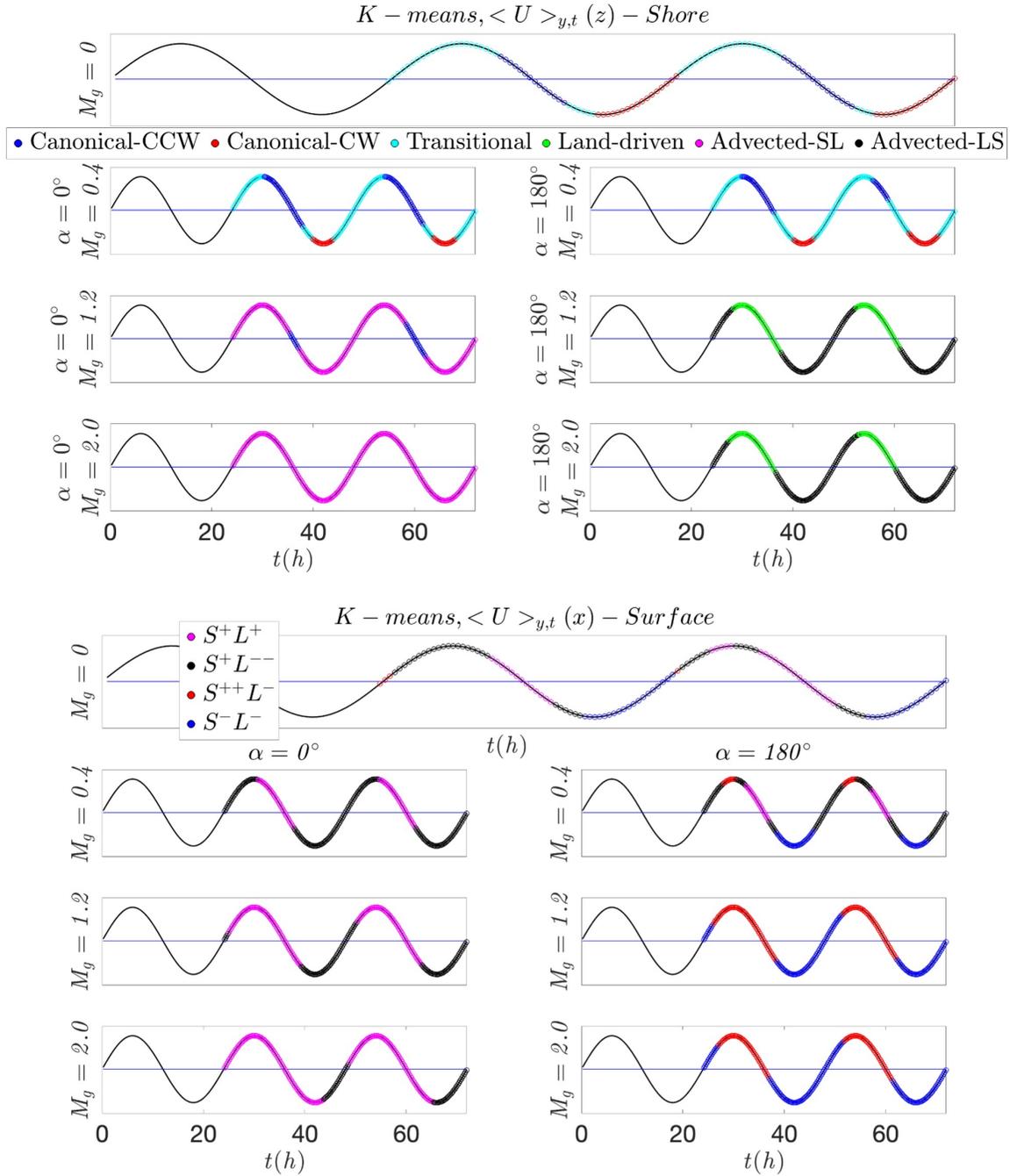

Fig. 10 Macro classification of all the transient simulations using the *k*-means algorithm with $U(z)_{shore}$ (top). Micro classification with $U(x)$ - surface (bottom)